\documentclass[12pt]{article}
\usepackage{epsfig}

\let\bar=\overbar

\def\Dslash{\not{\hbox{\kern-4pt $D$}}}
\def\dslash{\not{\hbox{\kern-2pt $\del$}}}

\def\msb{{\bar{\ssstyle M \kern -1pt S}}}

\def\Title#1{\begin{center} {\Large {\bf #1} } \end{center}}

\begin{document}

\Title{Three-body charmless $B\to Khh$ decays at Belle}

\bigskip\bigskip


\begin{raggedright}  

{\it Alexei Garmash (representing the Belle Collaboration) \index{Garmash, A.}\\
High Energy Accelerator Research Organization (KEK)\\
1-1 Oho, Tsukuba, JAPAN}
\bigskip\bigskip
\end{raggedright}


   Study of three-body $B$ decays can significantly broaden the understanding of 
$B$ meson decay mechanisms and provide additional possibilities for CP violation
searches. The decays of $B$ mesons to charmless three-body $Khh$ final states 
can be described by a $b\to u$ tree-level spectator diagram and a $b\to s(d)g$ 
one-loop penguin diagram.
(Although $b\to u$ $W$-exchange, annihilation, or vertical $W$ loop diagrams 
can also contribute to these final states, they are expected to be smaller and we
neglect them for simplicity.)
 Three-body decays of $B$ mesons to final 
states with odd numbers of kaons ($s$-quarks) are expected to proceed dominantly
via the $b\to sg$ penguin transition since the $b\to u$
contribution in these cases has an additional CKM suppression. In contrast, 
decays with two kaons in final state 
proceed via $b\to u$ tree transition (with possible contribution from 
$b\to dg$ penguin), and have no contribution from $b\to sg$ penguin.
From this simple consideration we can naively expect that the $K\pi\pi$ and $KKK$ 
final states have the largest signal among other three-body $Khh$ final states.
The observation of $K^{+(0)}\pi^+\pi^-$ and $K^{+(0)}K^+K^-$ final states has 
been reported recently by the Belle Collaboration~\cite{Abe:2002av,kshh}.
Here, we report the results of a study of $B$ mesons decays to three-body 
$K\pi\pi$, $KK\pi$, and $KKK$ final states, where no prior assumptions
are made about intermediate hadronic resonances. The inclusion of charge 
conjugate states is implicit throughout this report.

   The data sample used for this analysis was collected with the Belle 
detector~\cite{Belle} operating at the KEKB asymmetric energy $e^+e^-$ 
collider. 
It consists of 43~fb$^{-1}$ taken at the $\Upsilon(4S)$ 
resonance, corresponding to $45.3\times10^{6}$ produced $B\bar{B}$ pairs.

  Charged tracks are selected with a set of track quality 
requirements based on the average hit residual and on the distances 
of closest approach to the interaction point.
We also require that the transverse track momenta be greater 
than 0.1~GeV/$c$ to reduce the low momentum combinatoric background.
  Charged kaons and pions are identified basing on $dE/dx$ measurements by
the central drift chamber, information from threshold \v{C}erenkov counters
and time-of-flight scintillation counters. Tracks that identified as protons 
or electrons are rejected.
  Neutral kaons are reconstructed via their decay chain 
$K^0(\bar{K}^0)\to K_S\to\pi^+\pi^-$. The invariant mass of the two pions is 
required to satisfy $|M(\pi^+\pi^-)-M_{K^0}|<10$~MeV/$c^2$ and the displacement 
of the $\pi^+\pi^-$ vertex from the interaction point in the transverse 
($r$-$\phi$) plane is required to be greater than 0.1~cm and less than 20~cm.
The $K_S$ flight direction and combined pion pair momentum direction in the 
$r$-$\phi$ plane must agree within 0.2 radians. Photons are identified as
isolated clusters in the electromagnetic calorimeter with energy greater than
50~MeV. Pairs of photons with an invariant mass within 15~MeV of the $\pi^0$
nominal mass are considered as $\pi^0$ candidates. The reconstructed $\pi^0$ 
momentum is required to be more than 0.2~GeV/$c$.

  We reconstruct $B$ mesons in three-body $K\pi\pi$, $KK\pi$, and $KKK$
final states with charged or neutral (not more than one) pions and charged or
neutral kaons. The candidate events are identified by their center of mass (CM)
energy difference, $\Delta E=(\sum_iE_i)-E_{\rm b}$, and the beam constrained
mass, $M_{\rm bc}=\sqrt{E^2_{\rm b}-(\sum_i\vec{p}_i)^2}$, where 
$E_{\rm b}=\sqrt{s}/2$ is the beam energy in the CM frame, and $\vec{p}_i$ and
$E_i$ are the CM three-momenta and energies of the candidate $B$ meson decay
products. We select events with $M_{\rm bc}>5.20$~GeV/$c^2$ and 
$-0.30<\Delta E<0.50$~GeV, and define a {\it signal} region of 
$|M_{\rm bc}-M_B|<9$~MeV/$c^2$ and $|\Delta E|<0.04$~GeV and two $\Delta E$ 
{\it sideband} regions defined as $-0.08$~GeV $<\Delta E<-0.05$~GeV and 
$0.05$~GeV $<\Delta E<0.15$~GeV.

\begin{figure}[t]
\begin{center}
\hspace*{-1mm}\epsfig{file=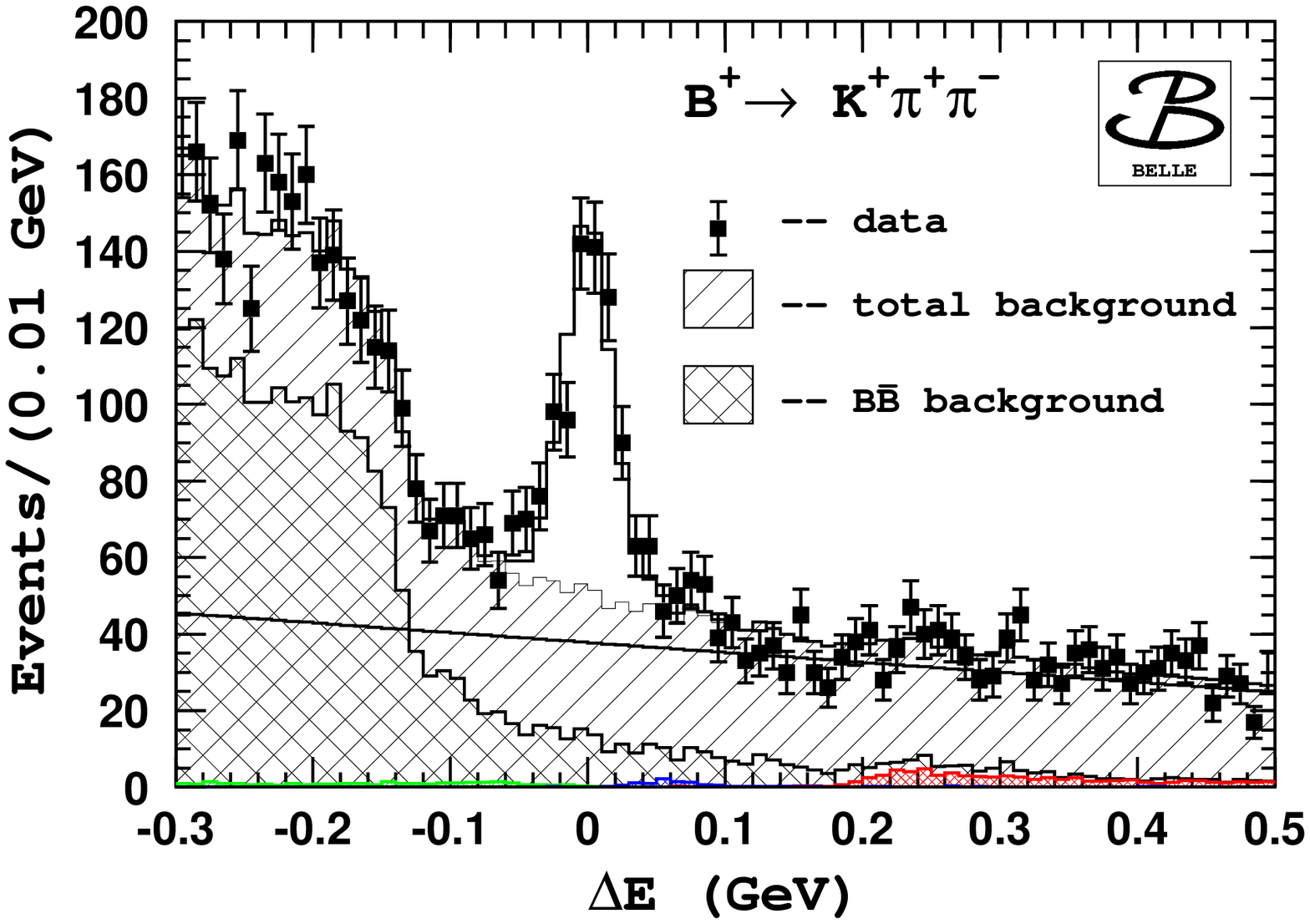,width=52mm,height=38mm}\hfill
\hspace*{-4mm}\epsfig{file=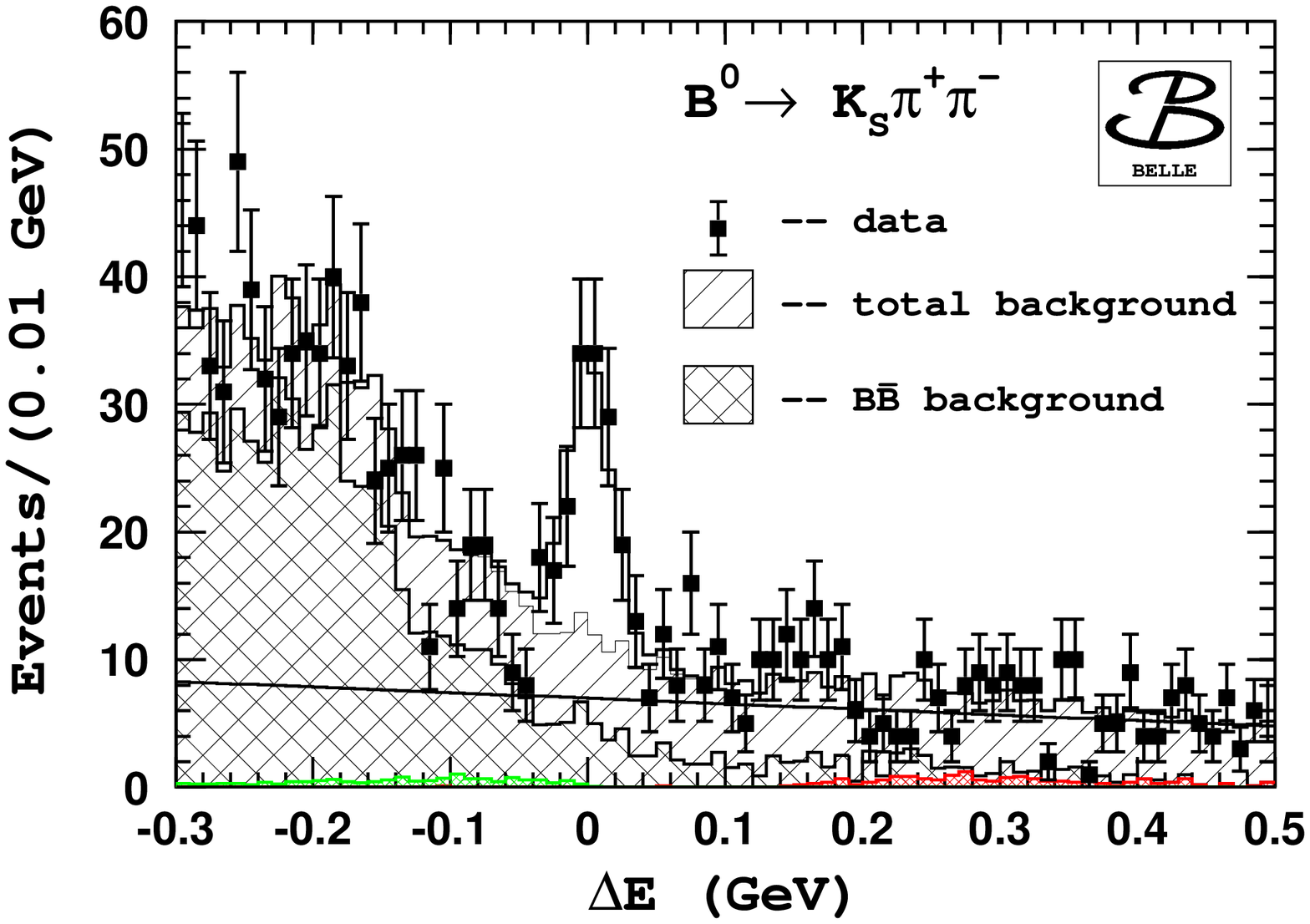,width=52mm,height=38mm}\hfill
\hspace*{-4mm}\epsfig{file=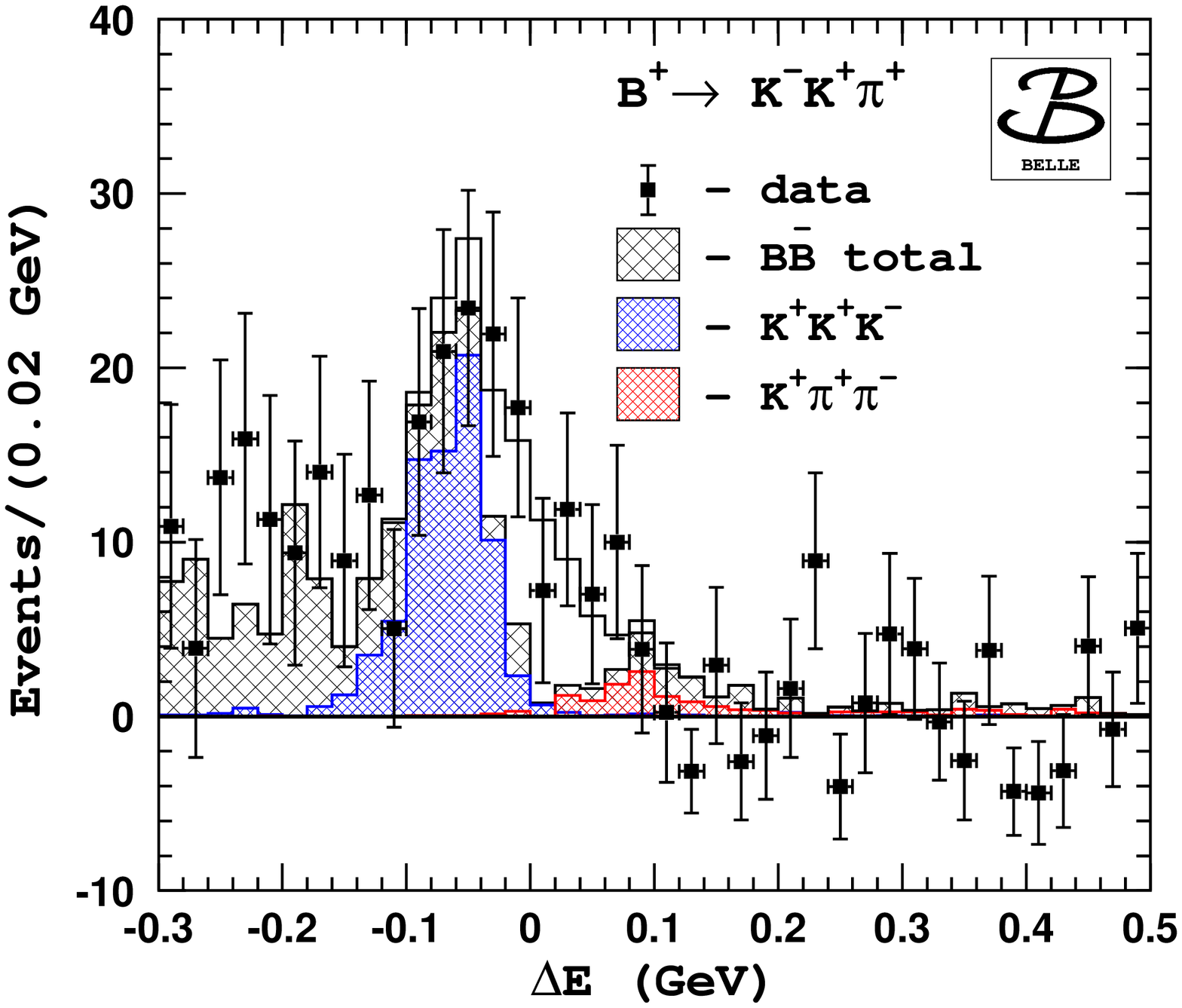,width=52mm,height=37mm}
\vspace*{-3mm}\\
\hspace*{-1mm}\epsfig{file=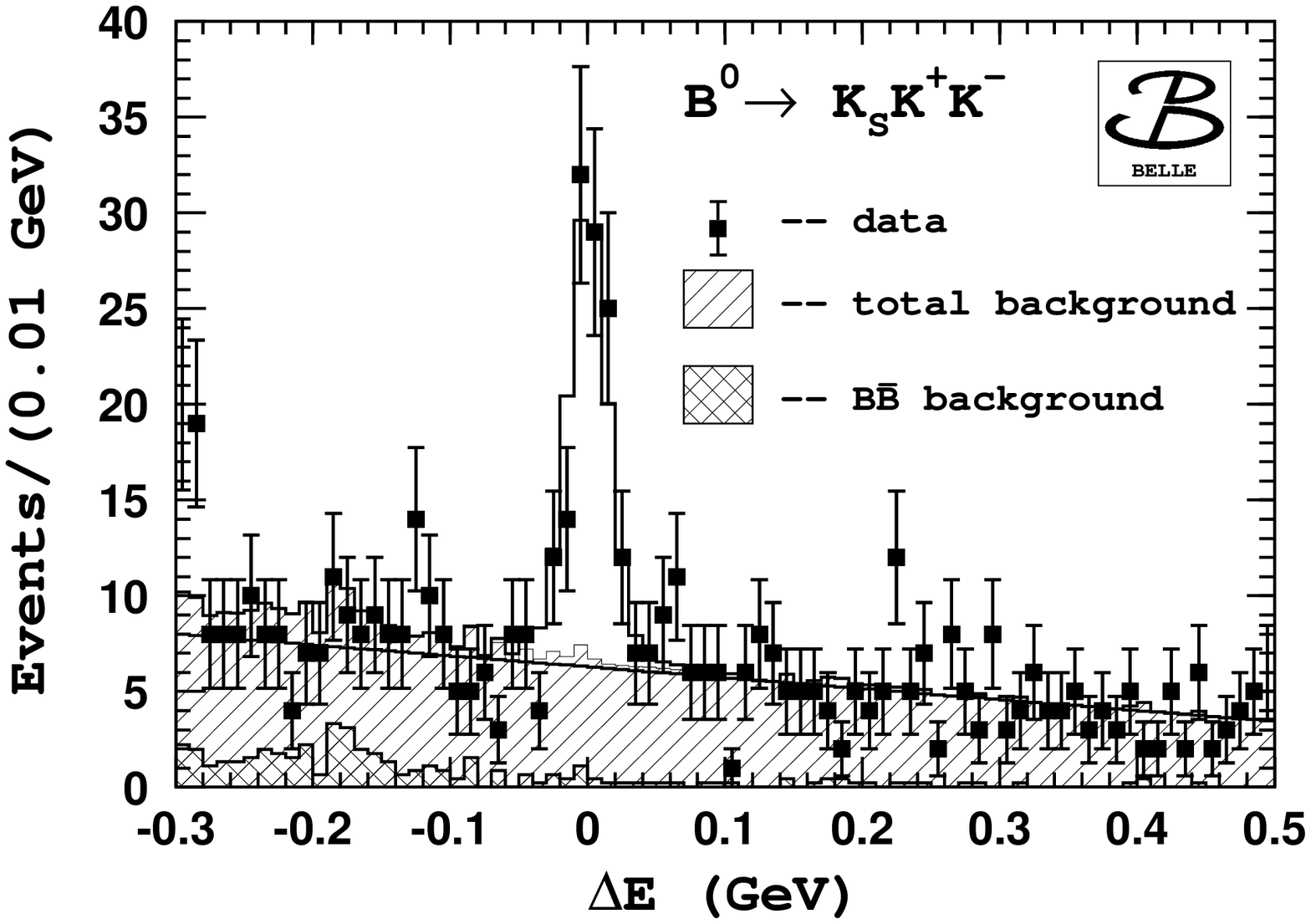,width=52mm,height=38mm}\hfill
\hspace*{-4mm}\epsfig{file=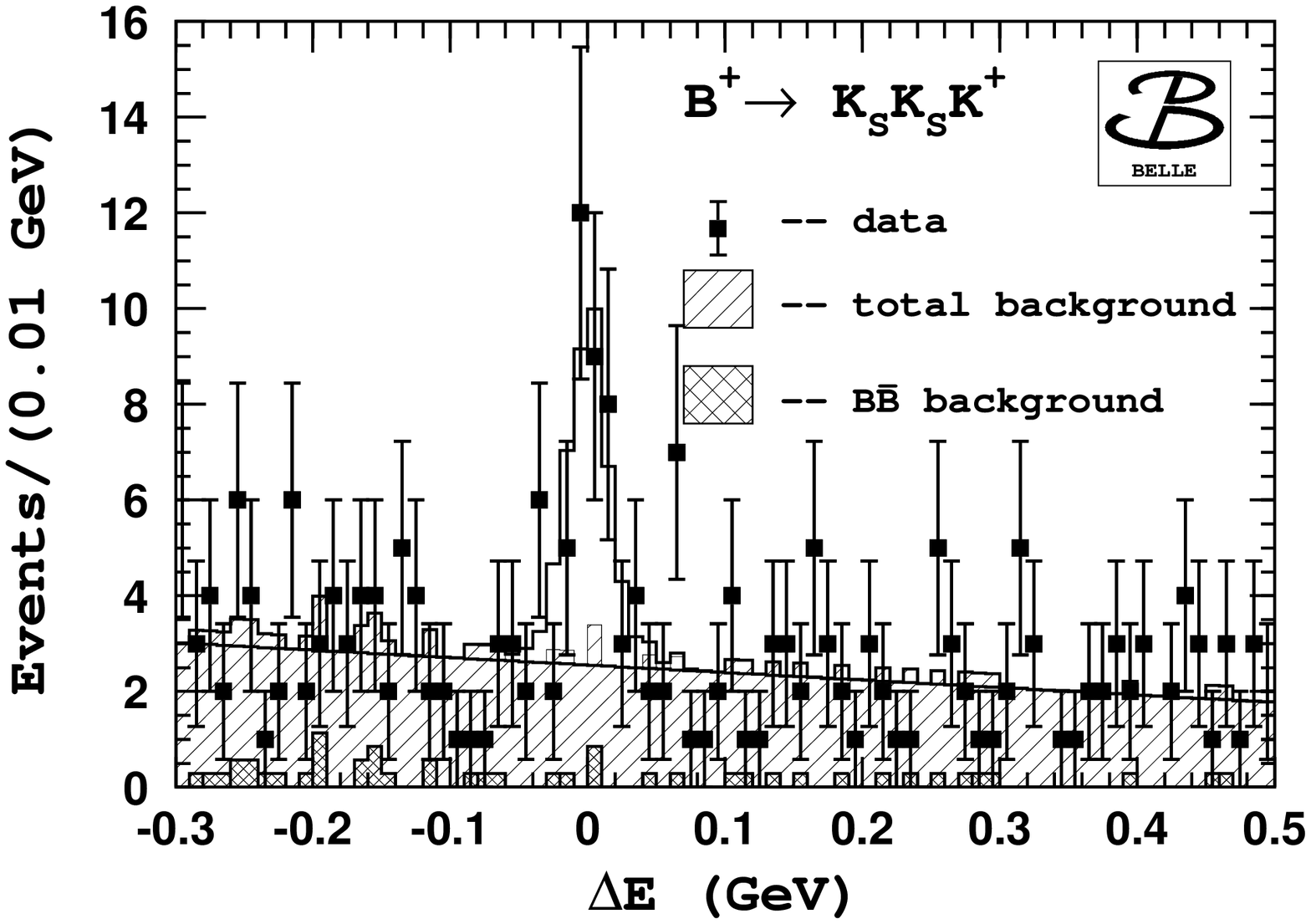,width=52mm,height=38mm}\hfill
\hspace*{-4mm}\epsfig{file=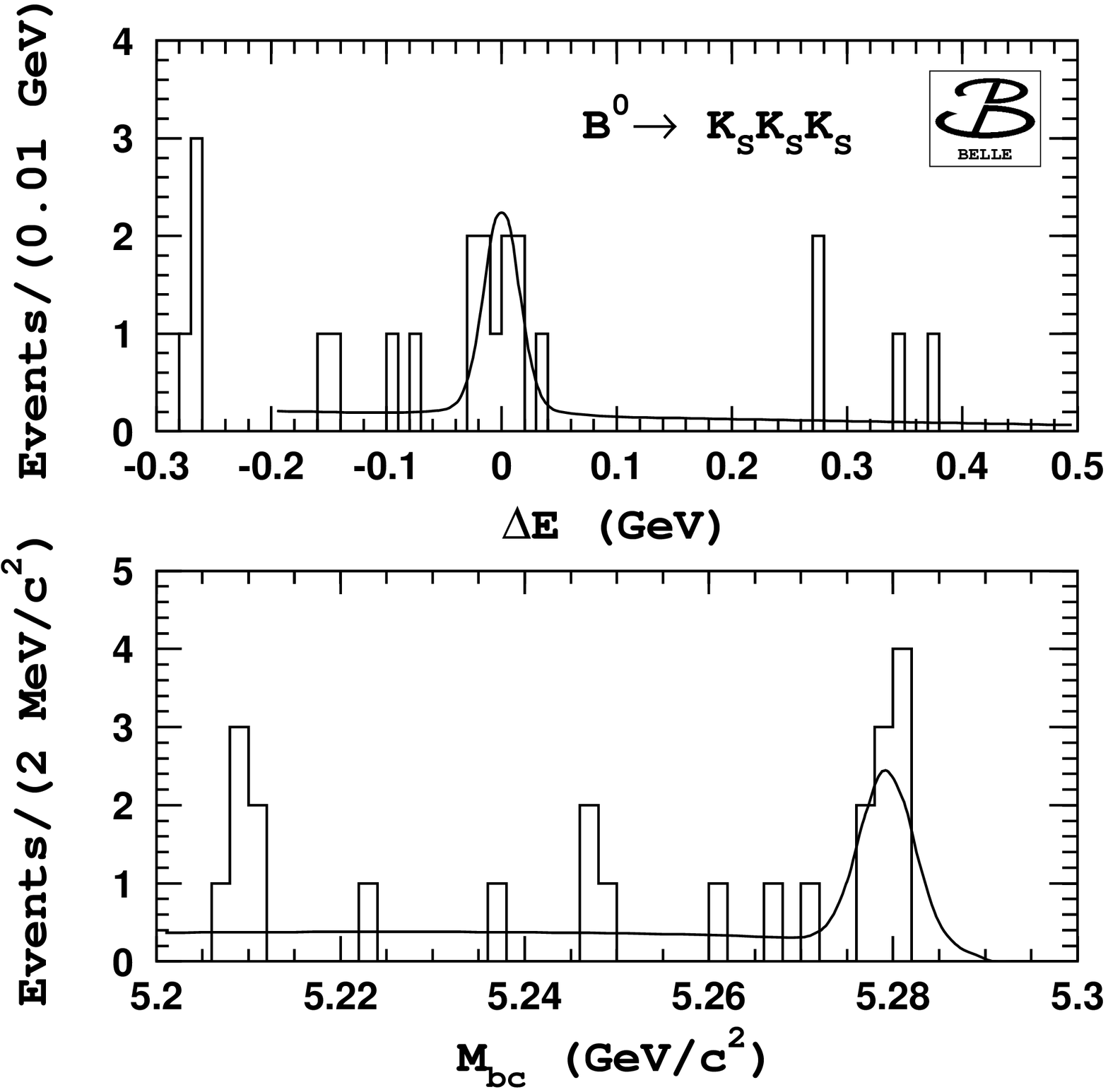,width=52mm,height=38mm}
\vspace*{-4mm}\\
\caption{The $\Delta E$ distributions for three-body final states.}
\label{fig:dE}
\end{center}
\end{figure}

   To suppress the large combinatorial background which is dominated by 
two-jet-like $e^+e^-\to~q\bar{q}$ continuum events, variables that characterize 
the event topology are used.
The procedure is described in detail 
in Ref.~\cite{Abe:2002av}.
The  potentially dangerous sources of background from other $B$ decays were
studied with a large sample of $B\bar{B}$ Monte Carlo (MC) events.
As a result of this study we found that $B\to Dh$, where $h$ stands for a charged 
pion or kaon, produce the dominant background to the most of the three-body final 
states. To suppress this
background we introduce $D$ veto cut: we reject candidates if any two-particle 
invariant mass is consistent with a $D\to K\pi$ hypothesis, independently of the
PID information. In the analysis of $K^{+(0)}\pi^+\pi^-$ final states we also
apply a charmonium veto: candidates, with a two-pion invariant mass consistent
with $J/\psi(\psi(2S))\to\mu^+\mu^-$ are rejected.

  The $\Delta E$ distributions for some of the three-body final states are 
shown in Fig.~\ref{fig:dE}, where the points with errors represent the 
experimental data, and histograms show the expected background distributions.
To extract the signal yield in the $K\pi\pi$ and $KKK$ final states, we fit the 
$\Delta E$ distributions to the sum of signal and background components. The 
signal shape is parameterized by a double Gaussian function with the same mean.
There are two sources of background: $q\bar{q}$ continuum background that is 
approximated by the linear function and $B\bar{B}$ background which is final 
state dependent. The shape of the $B\bar{B}$ background for each particular 
three-body final state was determined using a large set ($\sim$3.5 times the data
set) of MC events. The $q\bar{q}$ and $B\bar{B}$ component are also shown in 
Fig.~\ref{fig:dE}. The $B\bar{B}$ is found to be substantial in $K\pi\pi$ final
states, while it gives very small contribution in three-kaon final states. 
For other final states, a different technique is used for signal yield extraction.
We subdivide the $\Delta E$ region into 20~MeV bins and determine the signal yield
in each bin from the fit to the corresponding $M_{\rm bc}$ spectrum. The signal
yield from the $M_{\rm bc}$ fit is then plotted as a function of $\Delta E$.
The results for the $K^+K^-\pi^+$ final state is shown in Fig.~\ref{fig:dE} as
points with errors, where the $B\bar{B}$ contribution is also shown. The more 
detailed description can be found in Ref.~\cite{Abe:2002av}.

\begin{table}[t]
\begin{center}
  \begin{tabular}{lcccr}  \hline \hline
  ~Mode~   & Eff., \%  &
               Yield, events~   &
               ${\cal{B}}, 10^{-6}$ (43$\tt fb^{-1}$) &
               ${\cal{B}}, 10^{-6}$ (29$\tt fb^{-1}$), Ref.[1] \\ \hline \hline
  $K^+\pi^-\pi^+$  & $21.1$ &     $463\pm32$         & $59.3\pm4.1$ & $55.6\pm5.8\pm7.7$  \\
  $K^0\pi^-\pi^+$  & $5.23$ &    $94.7\pm14.4$       & $41.7\pm7.2$ & $53.2\pm11.3\pm9.7$ \\

  $K^+\pi^-\pi^0$  & $11.6$ & $173^{+30.5}_{-29.6}$  &  $-$  & $47.1\pm8.2\pm6.3$ \\

  $K^+K^+K^-$      & $22.2$ &     $289\pm20$         & $35.8\pm2.5$ & $35.3\pm3.7\pm4.3$  \\
  $K^0K^+K^-$      & $7.10$ &    $88.8\pm11.8$       & $32.3\pm4.8$ & $34.8\pm6.7\pm6.5$  \\
  $K_SK_SK^+$      & $5.76$ &     $27.5\pm6.7$       & $13.1\pm3.2$ & -- \\
  $K_SK_SK_S$      & $3.86$ &  $8.2^{+3.5}_{-2.9}$   & $5.5^{+2.3}_{-1.9}$ & -- \\
  $K^+K^-\pi^+$    & $13.8$ &     $49\pm15$          & $9.1\pm2.8 (<14)$ & $<12$\\
  $K^+K^+\pi^-$    & $14.2$ &     $-4.7\pm9$         & $<2.0$ & $<3.2$  \\
  $K^-\pi^+\pi^+$  & $17.0$ &     $14\pm12$          & $<5.4$ & $<7.0$  \\
  $K^0K^{\pm}\pi^{\mp}$ & $4.53$ & $1\pm11$          & $<9.2$ & $<13.4$ \\
  $K_SK_S\pi^+$    & $5.31$ &     $-6.4\pm8.1$       & $<3.3$ & -- \\ \hline \hline
  \end{tabular}
\caption{Summary table of three-body results.}
\label{tab:results}
\end{center}
\end{table}

  The results of the fit are summarized in Table~\ref{tab:results}. 
Statistically significant signals are observed in $K^{+(0)}\pi^+\pi^-$, 
$K^+\pi^-\pi^0$ and all three-kaon final states.
The resulting branching fractions for the three-body final states 
are presented in Table~\ref{tab:results}, where the results of 
Ref.~\cite{Abe:2002av} are also included for comparison.

\begin{figure}[t]
\begin{center}
\hspace*{-0mm}\epsfig{file=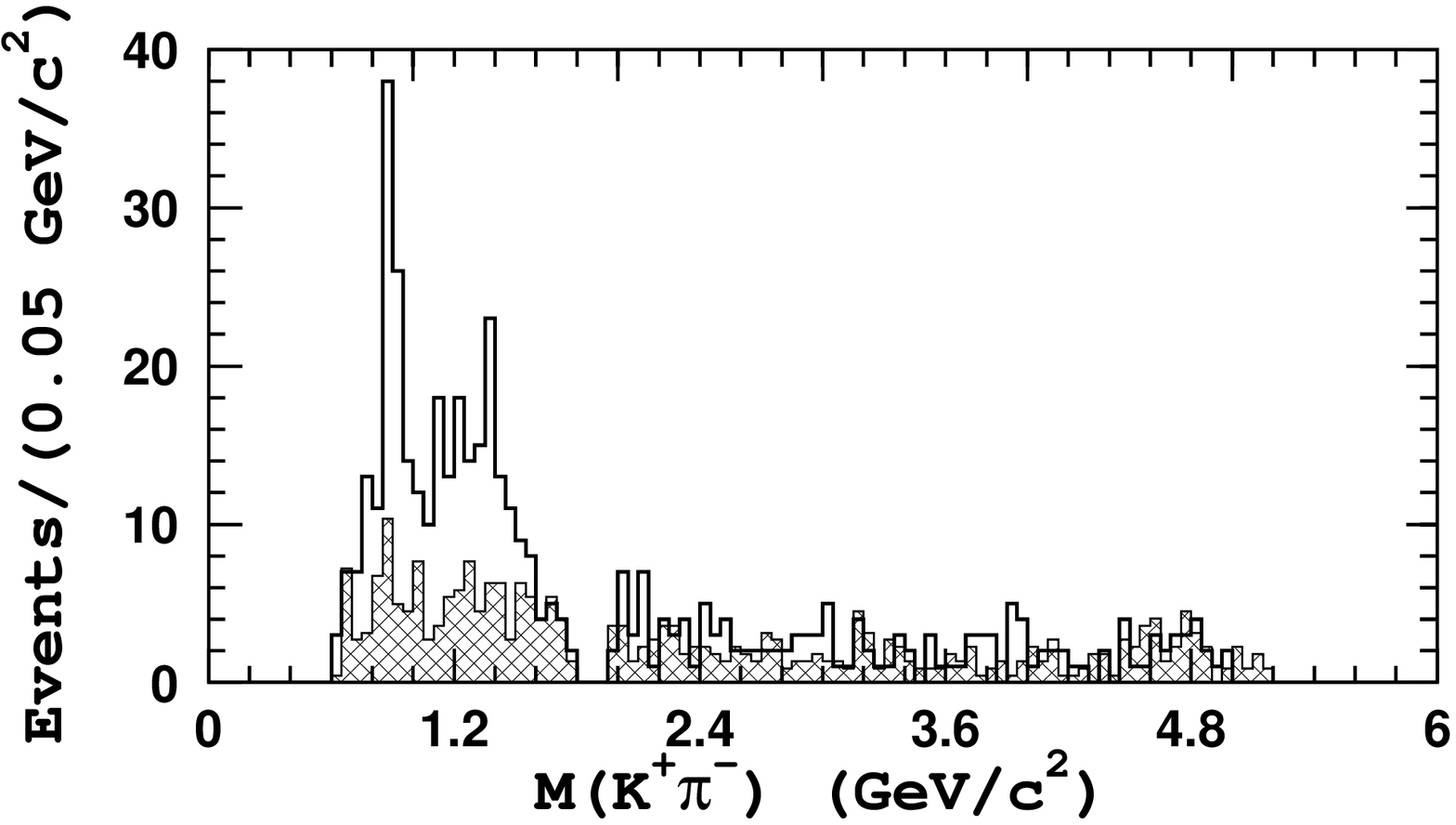,width=75mm,height=40mm}\hfill
\hspace*{-0mm}\epsfig{file=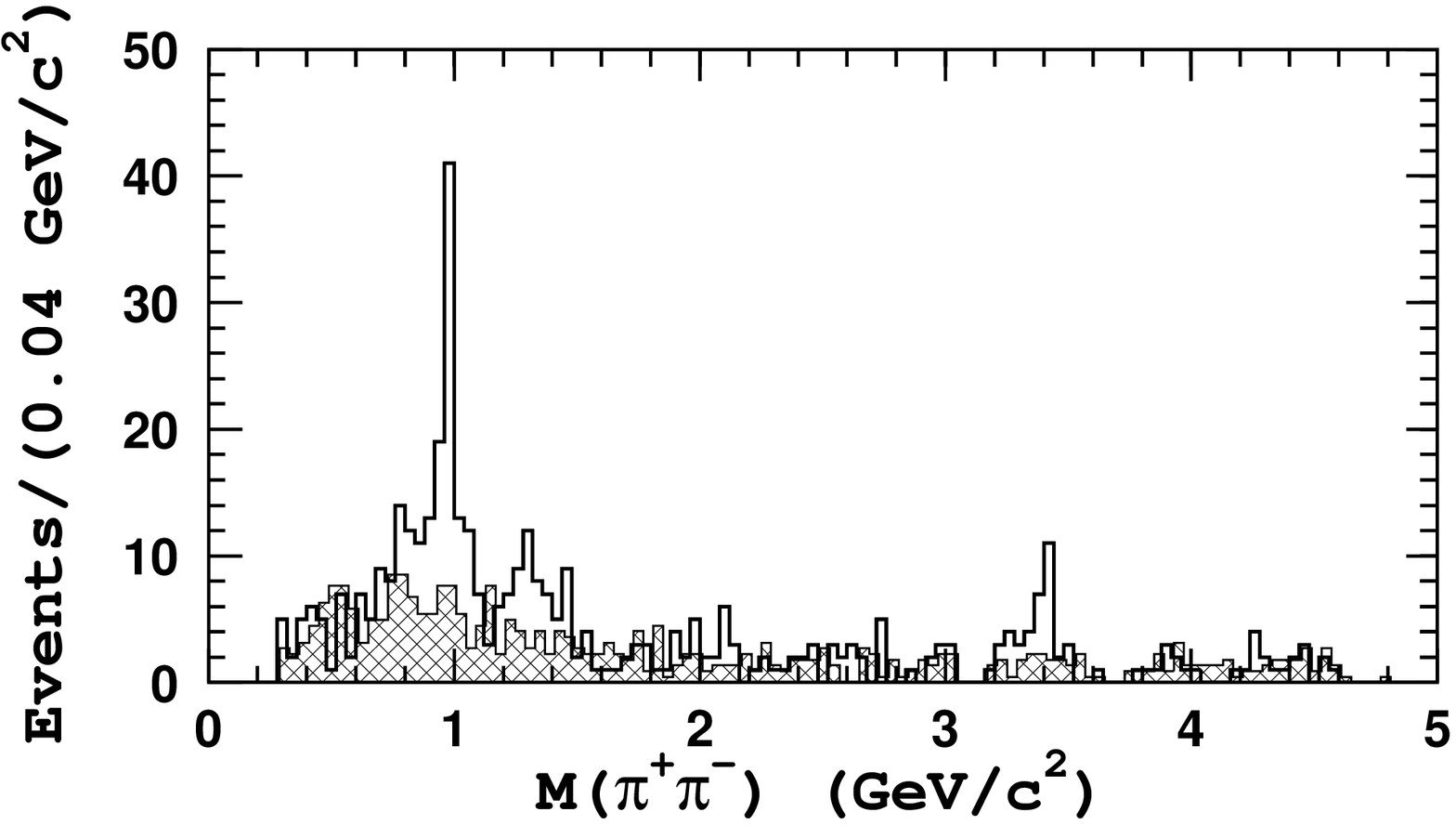,width=75mm,height=40mm} 
\vspace*{-4mm}\\
\caption{The $K^+\pi^-$ and $\pi^+\pi^-$ invariant mass spectra for 
         the $K^+\pi^+\pi^-$ final state.}
\label{fig:DP}
\end{center}
\end{figure}

To examine possible intermediate quasi-two-body states, we analyze two-particle
mass spectra. The $K^+\pi^-$ and $\pi^+\pi^-$ invariant mass spectra for the
$B^+\to K^+\pi^+\pi^-$ signal are shown in Fig.~\ref{fig:DP}.
To suppress the feed-across 
between the $\pi^+\pi^-$ and $K^+\pi^-$ states, we require the $K^+\pi^-$ 
($\pi^+\pi^-$) invariant mass to be larger than 2.0 (1.5)~GeV/$c^2$ when making
the $\pi^+\pi^-$ ($K^+\pi^-$) projection. The hatched histograms shown in 
Fig.~\ref{fig:DP} are the corresponding two-particle invariant mass spectra for
the background events in the $\Delta E$ sidebands with proper normalization.
The Dalitz plot analysis procedure is described in Ref.~\cite{Abe:2002av} in
detail.

  In conclusion, the results of the branching ratio measurement for the $B$
decays to three-body $K\pi\pi$, $KK\pi$, and $KKK$ final states are presented,
where the $K_SK_SK^+$ and $K_SK_SK_S$ final states are observed for the first
time. We also report 3$\sigma$ evidence for the signal in $K^+K^-\pi^-$ final
state. The analysis of quasi-two-body final states reveals large signals of
$B^+ \to f_0(980) K^+$, $B^+\to K^*(892)^0\pi^+$ in $K^+\pi^+\pi^-$ final 
state and $B^+\to \phi K^+$ in $K^+K^+K^-$ final state. The measured branching
fraction product for the $f_0(980) K^+$ final state is 
${\cal{B}}(B^+\to f_0(980)K^+)\times{\cal{B}}(f_0(980)\to \pi^+\pi^-)=
(9.6^{+2.5+1.5+3.4}_{-2.3-1.5-0.8})\times10^{-6}$. This is the first 
observation of a $B$ meson decay to a charmless scalar-pseudoscalar final state. 
  We find that effects of interference between different quasi-two-body 
intermediate states can have significant influence on the observed two-particle
mass spectra and a full amplitude analysis of three-body $B$ meson decays is 
required for a more complete understanding. This will be possible with 
increased statistics.

  We wish to thank the KEKB accelerator group for the excellent operation of 
the KEKB accelerator.

\end{document}